\documentclass[epj]{svjour}
\usepackage{amssymb}
\usepackage{graphicx}
\begin{document}
\title{Thermal conductivity of doped $\rm\bf La_2CuO_4$ as an example for heat transport by optical phonons in complex materials}
\author{C. Hess\inst{1} \and B. B\"{u}chner\inst{2}\inst{3}
}                     
%
%
\institute{D\'{e}partement de Physique de la Mati\`{e}re Condens\'{e}e, Universit\'{e} de Gen\`{e}ve, CH-1211 Gen\`{e}ve,  Switzerland \and 2. Physikalisches Institut, RWTH-Aachen, D-52056 Aachen, Germany \and Leibniz-Institute for Solide State and Materials Research, IFW-Dresden, 01171 Dresden, Germany}
\date{Received: date / Revised version: date}
%
\abstract{We investigate the phonon thermal conductivity $\kappa_{\mathrm{ph}}$ of doped $\rm La_2CuO_4$ based on out-of-plane thermal conductivity measurements. When room temperature is approached the temperature dependence of $\kappa_{\mathrm{ph}}$ strongly deviates from the $T^{-1}$-decrease which is usually expected for heat transport by acoustic phonons. Instead, $\kappa_{\mathrm{ph}}$ decreases much weaker or even increases with rising temperature. Simple arguments suggest that such unusual temperature dependencies of $\kappa_{\mathrm{ph}}$ are caused by heat transport via dispersive optical phonons.
} 
%
\maketitle
The thermal conductivity $\kappa$ is an interesting tool in order to probe dissipation and scattering of any propagating excitation in a solid. A recent example is $\kappa$ of complex materials with low-dimensional spin structures where magnetic excitations provide an unusual transport channel of heat \cite{Sologubenko00,Hess01,Hess02,Hess03b,Hess03}. Such magnetic heat conduction can in general only be measured in parallel with the phononic heat transport of the underlying crystal lattice. A thorough understanding of the phonon thermal conductivity $\kappa_{\mathrm{ph}}$ is therefore essential to identify and separate a magnetic contribution $\kappa_{\mathrm{mag}}$.

In early experiments on magnetic materials clear deviations from a low-temperature scaling as $T^3$, which as $T\rightarrow0$ is expected for $\kappa_{\mathrm{ph}}$, were assessed as one important indication of $\kappa_{\mathrm{mag}}$ \cite{Luethi62,Douglass63,Coenen77}. Recently, experiments have been performed on materials where magnetic coupling and velocities of magnetic excitations are several orders of magnitude larger than in these early studies. Significant $\kappa_{\mathrm{mag}}$ is in these cases present at much higher temperatures where $\kappa_{\mathrm{ph}}$ is expected to follow a $T^{-1}$ behavior. Prominent examples are the spin-ladder system $\rm (Sr,Ca,La)_{14}Cu_{24}O_{41}$ and the two-dimensional antiferromagnetic $\rm La_2CuO_4$, where pronounced peaks are observed in $\kappa$ at high $T$ \cite{Sologubenko00,Hess01,Hess02,Hess03b,Hess03}. These high-$T$ anomalies reflect the dimensionality of the underlying magnetic structure, i.e., they are only observed when $\kappa$ is measured along the direction of ladders or, in the case of $\rm La_2CuO_4$, parallel to the magnetic planes. However, it is not a priori clear that such anomalies result from magnetic heat transport. Just as well, it is possible that dispersive optical modes provide a thermal transport channel which could generate an unusual high-$T$ behavior of $\kappa_{\mathrm{ph}}$. Even the observed anisotropies could be explained in such a scenario, since low-dimensional magnetic structures often originate from lattice sub-structures with a similar low dimensionality. It is therefore essential to carefully investigate the origin of high $T$ anomalies. A firm proof that indeed magnetic heat conduction is present, is for example obtained if the analysis of the additional contributions yields characteristic magnetic properties like the spin gap of the ladders or magnetic defect distances \cite{Sologubenko00,Hess01,Hess03}.

The actual relevance of heat conduction by optical phonons has, however, never been studied systematically. It is the purpose of this paper to initiate future work in this field by pointing out that optical phonons provide a substantial transport mechanism of heat in doped $\rm La_2CuO_4$. Our estimation for the heat conductivity by optical phonons $\kappa_{\mathrm{opt}}$ is based on realistic values for the phonon energies and velocities in this material. We find that at room temperature $\kappa_{\mathrm{opt}}$ can amount to about 40\% of the total phonon thermal conductivity $\kappa_{\mathrm{ph}}$.

\begin{figure}
\includegraphics [width=\columnwidth,clip] {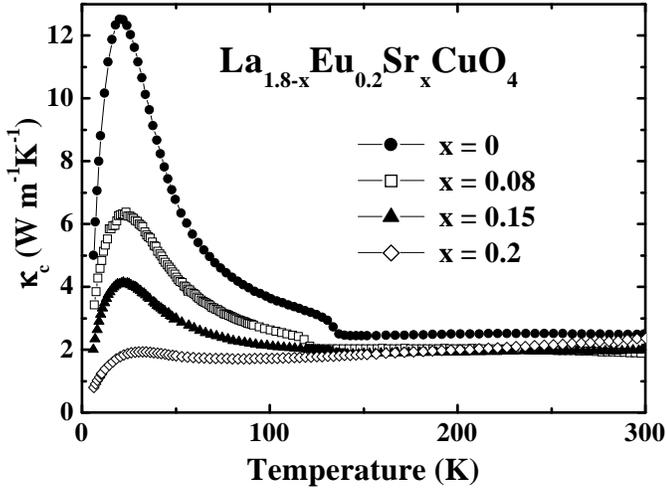}
\caption{\label{fig1}Out-of-plane thermal conductivity $\kappa_c$ of $\rm La_{1.8-x}Eu_{0.2}Sr_xCuO_4$ ($x=0$, 0.08, 0.15, 0.2) as a function of temperature.}
\end{figure}
\begin{figure}
\includegraphics [width=\columnwidth,clip] {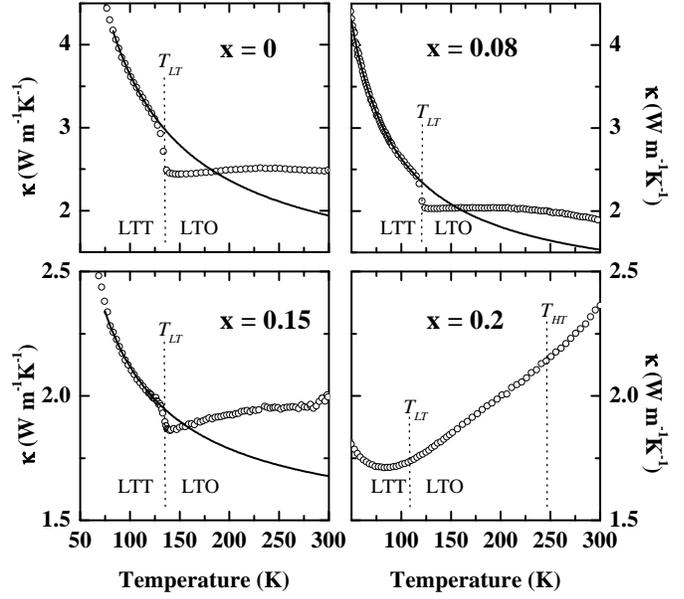}
\caption{\label{fig2}$\kappa_c$ of $\rm La_{1.8-x}Eu_{0.2}Sr_xCuO_4$ ($x=0$, 0.08, 0.15, 0.2) as a function of temperature (open circles). The solid lines extrapolate the $T$-dependence of $\kappa_c$ in the LTT-phase to high $T$ (cf. text).}
\end{figure}

The experimental data which give reason for our considerations is the out-of-plane thermal conductivity $\kappa_c$ of $\rm La_{1.8-x}Eu_{0.2}Sr_xCuO_{4}$-single crystals with $x=0$, 0.08, 0.15 and 0.2. Fig.~\ref{fig1} displays $\kappa_c$ as a function of $T$.
Prior to discussing the high-$T$ behavior of $\kappa_c$ in more detail, we briefly summarize the most important facets of the doping- and temperature dependence of $\kappa_c$ which have been elaborately discussed in Ref.~\cite{Hess03a}. Details of the experiment are also described there. Unlike the in-plane-thermal conductivity $\kappa_{ab}$ of doped $\rm La_2CuO_4$ which, depending on doping, consists of phononic, magnetic and electronic contributions, $\kappa_c$ is purely phononic for all Sr-contents. This is concluded from the very low out-of-plane electrical conductivity and the negligible magnetic coupling along the $c$-axis \cite{Hess03,Hess03a,Thio88,Nakamura91}. A usual low-$T$ phonon peak is present in $\kappa_c$ for all doping levels. The gradual suppression of the peak upon Sr-doping is straightforwardly understood in terms of scattering by impurities which are induced by the Sr-ions. For $x\leq0.15$ a step-like decrease is present at $T_{LT}\approx130$~K which may be seen more clearly in Fig.~\ref{fig2}, where the high-$T$ behavior of $\kappa_c$ is shown separately for each compound. At $T_{LT}$ a structural phase transition occurs between the so-called LTO- (Low-Temperature-Orthorhombic, $T\geq T_{LT}$) and LTT-phases (Low-Temperature-Tetragonal, $T\leq T_{LT}$). Soft tilting modes of the $\rm CuO_6$-octahedra enhance the scattering of acoustic phonons in the LTO-phase, which generates the step at $T_{LT}$.

As revealed from Fig.~\ref{fig2}, the $T$-dependence of $\kappa_c$ in the LTO-phase apparently evolves from ($x=0$, 0.08) an almost constant behavior with a slight decrease at high $T$ into a clear increase with increasing $T$ ($x=0.15$, 0.2). This obviously conflicts with the expected decrease of $\kappa_{\mathrm{ph}}$ (ideally following $T^{-1}$). Strong deviations from the expected behavior could in principle arise from structural peculiarities of the LTO-phase. This can, however, be ruled out since in the case of $x=0.2$ the LTT-, the LTO- and in addition the so-called HTT-structures (High-Temperature-Tetragonal) are successively passed through with rising $T$ without any influence on the increase of $\kappa_c$. Furthermore, $\kappa_{\mathrm{ph}}$ of the LTO-phase should never be equal to and never exceed $\kappa_{\mathrm{ph}}$ of the LTT-phase, where additional phonon-scattering by the aforementioned soft tilting modes does not exist. Yet this is the case for $x=0.15$ and $x=0.2$. For $x=0$ and $x= 0.08 $, a similar situation is present concerning a hypothetical $\kappa_{\mathrm{ph}}$ of the LTT-phase which is extrapolated into the LTO-phase (solid lines in Fig.~\ref{fig2}). These extrapolations have been obtained by fitting $\kappa_{\mathrm{ph}}$ of the LTT-phase with $\kappa=a/T+b$, $b>0$ and extrapolating this function to high $T$. In all cases these extrapolations are considerably smaller than the measured $\kappa_c$ just slightly above $T_{LT}$.\footnote{Note that an intersection between the measured and extrapolated curves also arises for extrapolation functions which decrease much weaker.}
We therefore conclude that in addition to the usual heat conductivity by the acoustic phonons, $\kappa_{\mathrm{acst}}$, a further transport channel for heat is present.
Electronic and magnetic contribution to $\kappa$ can be excluded as already mentioned above. The additional heat transport must therefore arise due to optical phonons. This is remarkable since in practice optical phonons are neglected in the analysis of phonon heat transport.

It is commonly argued that optical phonons have dispersion curves at high energy with almost no slope. In that case the number of excited phonons is small and these phonons do not propagate. Their ability to transport heat is therefore negligibly small. The situation is different in the case of doped $\rm La_2CuO_4$ where it is well established that numerous optical phonons have a high velocity \cite{Pintschovius90,Pintschovius91}. Considerable transport of heat by such optical phonons should therefore be possible if their energy is not too high to be excited at temperatures relevant here, i.e. at $T\lesssim 300$~K. In the following we therefore investigate this possibility based on a simple kinetic model for the phononic heat transport.

Following standard linearized Boltzmann-theory the thermal conductivity of a single phonon branch (labelled by $i$) in a crystal with cubic symmetry is given by \cite{Callaway}
\begin{equation}
\label{ansatz}\varkappa^i=\frac{1}{3}\frac{1}{(2\pi)^3}\int v^i_{\bf k}l^i_{\bf k}\epsilon^i_{\bf k}\frac{d}{dT}n^i_{\bf k}d{\bf k},
\end{equation}
where $v^i_{\bf k}$, $l^i_{\bf k}$, $\epsilon^i_{\bf k}$ and $n^i_{\bf k}$ denote velocity, mean free path, energy and Bose function of a phonon. Frequently, a Debye-approximation with momentum independent mean free path is used, i.e. $v^i_{\bf k}\equiv v^i$ and $l^i_{\bf k}\equiv l^i$. In this case Eq.~\ref{ansatz} simplifies to
\begin{equation}
\label{estimate}\varkappa^i=\frac{1}{3}c^i_Vv^i\, l^i
\end{equation}
with the specific heat $c^i_V$ of that phonon branch. The usual acoustic phonon thermal conductivity $\kappa_{\mathrm{acst}}$ is then obtained by summing up the contributions of the three acoustic phonon branches. Often the different branches of acoustic phonons are unified by employing an avergaged phonon velocity and the Debye specific heat of phonons. Even though such an estimate for $\kappa_{\mathrm{acst}}$ is very crude, yet it is helpful to understand the $T$ dependence of  $\kappa_{\mathrm{acst}}$ in the low- and high-$T$ limits: At low $T$ the dominating scattering process is temperature independent scattering on crystal boundaries which leads to $\kappa_{\mathrm{acst}}\propto c_V \propto T^3$. For $T\rightarrow\infty$ the specific heat becomes constant while the phonon mean free path is inversely proportional to the number phonons that generate umklapp processes, i.e., $\kappa_{\mathrm{acst}} \propto T^{-1}$. These two limiting $T$-dependencies lead to the characteristic phonon-peak of $\kappa_{\mathrm{acst}}$ at low-$T$ which arises when umklapp scattering becomes important and over-compensates the increasing number of phonons.

In order to obtain a description which takes optical phonon branches into account we take a summation over {\em all} relevant phonon branches (acoustic and optic), i.e.,
\begin{equation}
\label{summi} \kappa=\sum_i\varkappa^i=\frac{1}{3}\sum_ic_V^iv^il^i,
\end{equation}
where we extract the velocities $v^i$ from linear approximations to the experimental ($0,0,q$)-phonon dispersion relations as measured by inelastic neutron diffraction \cite{Pintschovius91}. The specific heat $c_{V}^i$ of each branch is calculated with the usual Debye model where an energy gap $\Delta_{\mathrm{opt}}^i$ has to be considered for the optical branches. Hence we have
\begin{equation}
c_V^i=3N_Ak_B\left(\frac{T}{\Theta_D^i}\right)^3I^i,
\end{equation}
where $N_A$ and $k_B$ are Avogadro's and Boltzmann's constants, $\Theta_D^i=v^i\hbar k_B^{-1}(6\pi^2n)^{1/3}$ with $n$ as the density of unit cells and
\begin{equation}
I^i=\int_0^{\frac{\Theta_D^i}{T}}\left(x+\frac{\Delta_ {\mathrm{opt}}^i}{T}\right)^2
\frac{x^2\exp(x+\frac{\Delta_ {\mathrm{opt}}^i}{T})}{[\exp(x+\frac{\Delta_ {\mathrm{opt}}^i}{T})-1]^2}dx.
\end{equation}
The different $v^i$, $\Delta_ {\mathrm{opt}}^i$ and $\Theta^i_D$ of the phonon branches which we used for our analysis are summarized in Table~\ref{tab1}.
\begin{table}
\center{\begin{tabular}{llll}\hline\hline \multicolumn{1}{c}{Type}&\multicolumn{1}{c}{$\,\,\Delta_ {\mathrm{opt}}^i$~(K)$\,\,$}&\multicolumn{1}{c}{$\,\, v^i~\rm (ms^{-1})\,\,$}&\multicolumn{1}{c}{$\,\,\Theta_D^i$~(K)$\,\,$}\\
\hline
acoustic 1+2&\multicolumn{1}{c}{0}&\multicolumn{1}{c}{5423}&\multicolumn{1}{c}{280}\\
acoustic 3&\multicolumn{1}{c}{0}&\multicolumn{1}{c}{12289}&\multicolumn{1}{c}{635}\\
optic A&\multicolumn{1}{c}{251}&\multicolumn{1}{c}{4263}&\multicolumn{1}{c}{220}\\
optic B&\multicolumn{1}{c}{333}&\multicolumn{1}{c}{3108}&\multicolumn{1}{c}{161}\\
optic C&\multicolumn{1}{c}{359}&\multicolumn{1}{c}{28041}&\multicolumn{1}{c}{1449}\\
\hline\hline
\end{tabular}}
\caption{\label{tab1}Optical gaps $\Delta_ {\mathrm{opt}}^i$, velocities $v^i$ and 'Debye' temperatures $\Theta^i_D$ of the considered phonons.}
\end{table}

\begin{figure}
\includegraphics [width=.9\columnwidth,clip] {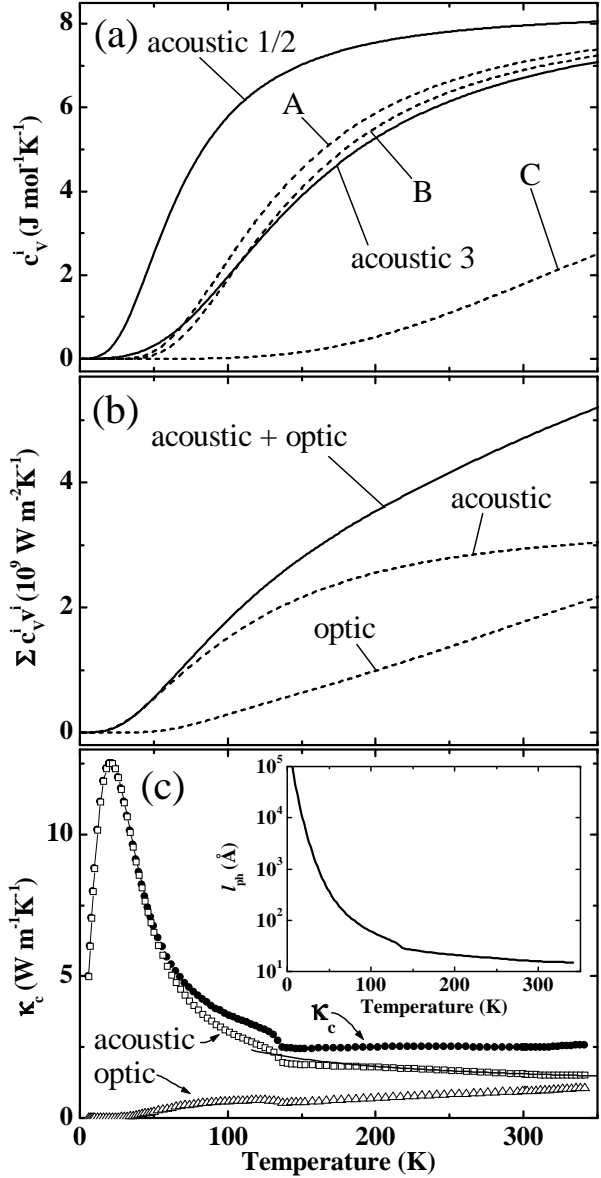}
\caption{\label{fig3}(a) Debye-specific heats $c_V^i$ of acoustic and optic phonon branches as a function of $T$. (b) Different summations (optical, acoustic and acoustic+optic) of the products $v^ic_V^i$ as a function of $T$. (c) $\kappa_c$ (full circles) of $\rm La_{1.8}Eu_{0.2}CuO_4$ and the calculated
contributions by acoustic acoustic (open squares) and optic (open triangles) contributions. The solid line illustrates the $T^{-1}$-dependence of $\kappa_{\mathrm{acst}}$ at high $T$. Inset:
$l_{\mathrm{ph}}$ of $\rm La_{1.8}Eu_{0.2}CuO_4$.}
\end{figure}

It is evident from the table that the $v^i$ of the optical and acoustic branches are of comparable magnitude and therefore a significant propagation of the optical modes is present indeed. The degree of excitation of the optical modes is reflected by the specific heat, which is shown in Fig.~\ref{fig3}~(a), for the different phonon branches. Remarkably, at $T=300$~K the $c^i_V$ of the optical branches labelled A and B already amount to about 80\% of the Dulong-Petit value, i.e., the major part of these optical modes is already populated at this $T$.\footnote{Since only a single phonon branch is considered, the Dulong-Petit value is $24.9/3~\rm Jm^{-1}K^{-1}=8.3~\rm Jm^{-1}K^{-1}$.} $c^i_V$ of the strongly dispersing branch C apparently evolves at much higher $T$.

In our kinetic model the products $c_V^iv^i$, which measure the excitation of {\em propagating}
phonons, are more relevant for the heat transport. In Fig.~\ref{fig3}~(b) the $T$-dependence of the sum ${\cal S}_{\mathrm{total}}=\sum_ic_V^iv^i$ of all contributions is shown
as a solid line while the different contributions ${\cal
S}_{\mathrm{acst}}=\sum_i^{\mathrm{acst}}c_V^iv^i$ and ${\cal
S}_{\mathrm{opt}}=\sum_i^{\mathrm{opt}}c_V^iv^i$ of acoustic and optical phonons are displayed as
broken lines. Apparently, propagating optical phonons become significantly excited at $T\gtrsim
70$~K and already contribute 40\% of ${\cal S}_{\mathrm{total}}$ at 300~K.

%

It is now very interesting to use this information and analyze experimental phononic thermal
conductivities in terms of acoustic and optic contributions, $\kappa_{\mathrm{acst}}$ and
$\kappa_{\mathrm{opt}}$. If we follow our model, an accurate separation of $\kappa_{\mathrm{acst}}$
and $\kappa_{\mathrm{opt}}$ is now possible with the separate phonon mean free paths $l^i$ of each
branch. These are, however, not known. As a first approximation we therefore assume an integrative mean
free path $l^i\equiv l_{\mathrm{ph}}$. The total phonon thermal conductivity then separates into
acoustic and optic contributions in the same ratio as ${\cal S}_{\mathrm{acst}}$ and ${\cal
S}_{\mathrm{opt}}$. As an example, we have calculated $\kappa_{\mathrm{acst}}$
and $\kappa_{\mathrm{opt}}$ of $\rm La_{1.8}Eu_{0.2}CuO_4$ as a function of $T$ as shown in Fig.~\ref{fig3}~(c).\footnote{As a first approximation we assume for the calculation of $\kappa_{\mathrm{acst}}$ and $\kappa_{\mathrm{opt}}$ that the structural phase transition at $T_{LT}\approx130~$K does not affect the lattice dynamics.} Apart from the remarkable high $\kappa_{\mathrm{opt}}$, the $T$-dependence of $\kappa_{\mathrm{acst}}$ at high $T$ is now indeed in accordance with the expected decrease with increasing $T$. In fact, at $T\gtrsim200$~K, $\kappa_{\mathrm{acst}}$ even follows a $T^{-1}$-dependence as is illustrated in Fig.~\ref{fig3}~(c) by the solid line which represents a fit $\kappa_{\mathrm{acst}}=a/T+b$, $b>0$.

The inset of Fig.~\ref{fig3} depicts the $T$-dependence of $l_{\mathrm{ph}}$ of $\rm La_{1.8}Eu_{0.2}CuO_4$ which we have calculated from our experimental data and ${\cal S}_{\mathrm{total}}$. Its strong decrease has to be attributed to the growing importance of umklapp scattering processes with increasing $T$. Note that $l_{\mathrm{ph}}\approx 15${~\AA} at $T\approx 350$~K. This value is slightly larger than the lattice constant $c=13.2$~{\AA} which defines the lower limit for $l_{\mathrm{ph}}$. At high $T$, our estimation of $\kappa_{\mathrm{opt}}$ is therefore close to a {\em lower limit} of $\kappa_{\mathrm{opt}}$, i.e., optical contributions could even be more important than it is revealed in our study.

Several simplifications in our approach should be mentioned besides the integrative $l_{\mathrm{ph}}$. First, the linear approximation of the dispersion curves neglects curvatures of the real dispersion relations which are present in some parts of the Brillouin-zone. Second, even though in the real material comparable dispersing optical phonon modes are present along {\em all} crystal directions, it is certainly not perfectly isotropic as implied by our approach. Third and finally, apart from the three acoustic branches we only take three optical branches into account. It is not clear whether the first two simplifications influence our estimation towards a smaller or larger result for $\kappa_{\mathrm{opt}}$. Yet it is unlikely that the order of magnitude of $\kappa_{\mathrm{opt}}$ is significantly smaller. Taking more optical branches into account, as required for a more realistic description, will certainly shift $\kappa_{\mathrm{opt}}$ towards higher values. 

Our data suggest a clear doping dependence of the ratio of $\kappa_{\mathrm{opt}}$ and $\kappa_{\mathrm{acst}}$ since the high-$T$ increase of $\kappa_c$ becomes more and more apparent with increasing Sr-content (cf. Fig.~\ref{fig2}). We note in this context that for some of the dispersive optical phonon branches the optical gaps and the slopes of the dispersion curves change upon Sr-doping. Therefore, a growing importance of $\kappa_{\mathrm{opt}}$ with increasing Sr-content could, for example, be related to decreasing optical gaps and increasing slopes of the dispersion curves \cite{Pintschovius90,Pintschovius91}. However, in order to verify such a detailed interpretation, a more accurate description of $\kappa_{\mathrm{ph}}$ than our simple approach is necessary. In principle, such can be achieved when the full set of $v^i_{\bf k}$, $\epsilon^i_{\bf k}$ and $l^i_{\bf k}=v^i_{\bf k}\tau^i_{\bf k}$ (with the phonon life time $\tau^i_{\bf k}$), accessible through neutron scattering experiments, is employed into the $\varkappa^i$ in Eq.~\ref{ansatz}. 

In conclusion, we have used a simple model in order to analyze the phononic out-of-plane thermal conductivity $\kappa_c$ of $\rm La_{1.8-x}Eu_{0.2}Sr_xCuO_{4}$ in terms of contributions by acoustic and optic phonons. Heat conductivity by optical phonons turns out to be significant in this material, since according to our approach it amounts at room temperature to about 40\% of the total phonon thermal conductivity. This result has important implications for studies on magnetic heat transport of low dimensional spin systems which are currently on debate \cite{Sologubenko00,Hess01,Hess02,Hess03b,Hess03,Sologubenko00a,Sologubenko01,Lorenz02}. Unusual $T$-dependencies of $\kappa$ which are hastily attributed to magnetic heat transport must carefully be excluded to stem from heat transport by dispersive optical phonons. We stress that more careful analysis of this usually underestimated mechanism of heat transport is desirable.

C.H. acknowledges support by the DFG through grant HE3439/3-1. We thank M. Braden for helpful discussions and N. Jenkins for proofreading the manuscript.

\end{document}